%Paper: cond-mat/9312075
%From: alex@hal.icqem.pi.cnr.it (Alessandro Fortunelli)
%Date: Mon, 20 Dec 1993 11:20:47 +0100

\rm
\overfullrule=0pc
\magnification=1050
\parindent=0.9truecm
\vsize=24.5truecm
\hsize=10truecm
\baselineskip=0.92truepc
\nopagenumbers

\def\do{doubly occupied}
\def\dob{doubly occupied }
\def\ha{Hamiltonian}
\def\hab{Hamiltonian }

\def\huh{Hubbard Hamiltonian}

\def\hum{Hubbard model}
\def\humb{Hubbard model }
\def\or{orbital}
\def\orb{orbital }
\def\re{relaxation}
\def\reb{relaxation }
\def\si{singly occupied}
\def\sib{singly occupied }
\def\tJ{$t-J$ }
\def\tJm{$t-J$ model}
\def\tJmb{$t-J$ model }
\def\tJt{$t-J-t^\prime$ }
\def\tJtm{$t-J-t^\prime$ model}
\def\tJtmb{$t-J-t^\prime$ model }

\def\tim{$t_1-t_2-t_3-U$ model}
\def\timb{$t_1-t_2-t_3-U$ model }

\vskip 2cm

\centerline{\bf The \orb \re : a possible origin of \tJmb
with large $J$}
\bigskip \bigskip \bigskip
\bigskip \centerline{\bf Alessandro Fortunelli$^1$
and Anna Painelli$^2$}
\bigskip
\centerline{\it $^1$Istituto di Chimica Quantistica ed Energetica
Molecolare del C.N.R.,}
\centerline{\it Via Risorgimento 35, 56126 Pisa, Italy }
\centerline{\it and}
\centerline{\it $^2$Istituto di Chimica Fisica, Universita' di Parma,}
\centerline{\it Viale delle Scienze, 43100 Parma, Italy. }

\vfill\eject
\centerline {\bf ABSTRACT} \par

Whereas the \tJmb has gained wide popularity among physicists
as a model for high-$T_C$ superconductivity, its physical origin
is not at all clear. In this communication we show that
the Hubbard model with occupation-dependent hopping
(\tim), recently proposed to account for the relaxation of
doubly-occupied orbitals, reduces, in a physically relevant
parameter region, to an effective \tJtmb with no upper bound on the
$J/t$ ratio. Results of exact diagonalization studies on finite
size systems demonstrate the equivalence between the \timb with
intermediate$/$large U and the \tJt even for very large $J/t$ ratio
(up to 16). On the contrary, $t-U$ and \tJm s turn out to be equivalent
to \tJtmb only for very small $J$ values ($J$ much smaller than $t$).
Implications of the results on high-$T_C$ superconductivity
are briefly discussed.

\vfil \eject

Since the pioneering suggestion by Anderson$^{1}$,
strongly interacting electrons are generally believed to be
the key to understand high-temperature superconductivity.
In this respect the \tJmb has gained wide popularity
among physicists.
The \tJm $^{2}$ is defined by the \ha :

$$
H_{tJ} = - t \sum_{<i,j>, \sigma}
( \tilde a^+_{i, \sigma} \tilde a_{j, \sigma} + {\rm h.c.}) +
J \sum_{<i,j>} ( {\bf S}_i \cdot {\bf S}_j - {n_i n_j \over 4})
\eqno(1)$$
where the brackets $<i,j>$ denote pairs of nearest neighbor sites,
$n_i$ is the number operator: $n_i = (a^+_{i, \alpha} a_{i, \alpha}
+ a^+_{i, \beta} a_{i, \beta})$, ${\bf S}_i$ are spin operators:
${\bf S}_i = (1/2) \sum_{\sigma,\sigma^\prime} \tilde a^+_{i, \sigma}
{\bf \sigma}_{\sigma,\sigma^\prime} \tilde a_{i, \sigma^\prime}$, and
$\tilde a^+_{i, \sigma}, \tilde a_{i, \sigma}$
are electron creation and annihilation operators which act in
the reduced Hilbert space without \dob sites, i.e.:

$$
\tilde a^+_{i, \sigma} = a^+_{i, \sigma} (1 - n_{i, - \sigma})
\quad\quad\quad
\tilde a_{i, \sigma} = a_{i, \sigma} (1 - n_{i, - \sigma})
\eqno(2)$$
The term proportional to $n_i n_j$ is often dropped.
The \hab (1) is very simple and, due to the reduced
configuration space it works in, it is particularly amenable
to numerical
treatment (see, e.g.,$^{3}$). It shows a very rich phase
diagram ranging from antiferromagnetism to phase segregation$^{4}$.
However the physical origin of the \tJm , or at
least of the large $J$ required to get the interesting physics,
is not at all clear$^{5}$.
The model itself is
not "clearly defined": the expansion of the \humb \hab
($t-U$ model):

$$
H_{Hub} = - t \sum_{<i,j>, \sigma}
( a^+_{i, \sigma} a_{j, \sigma} + {\rm h.c.}) +
U \sum_i n_{i, \alpha} n_{i, \beta}
\eqno(3)$$
in the strong coupling ($U>>t$) limit does not lead to the
\tJ but to the \tJt \ha $^{2,5,6}$:

$$\eqalign{
H_{tJt^\prime} = H_{tJ} +
t^\prime \sum_{<i,j,k>, \sigma} &
( \tilde a^+_{k, \sigma} \tilde n_{i, - \sigma} \tilde a_{i, \sigma} -
\cr &
\tilde a^+_{k, \sigma} a^+_{j, - \sigma} a_{j, \sigma}
\tilde a_{i, - \sigma} + {\rm h.c.})
}\eqno(4)$$
with $J = (4 t^2 / U)$ and $t^\prime = (t^2 / U)$,
where $<i,j,k>$ stands for three different sites, with $i,k$
nearest neighbors of $j$.
One gets a $J/t$ ratio: $J/t = (4 t / U) << 1$
by hypothesis.
In the limit $J << t$ (i.e.,
$U >> t$ for the \huh) the \tJtmb
reduces to the \tJm , but as we will show in the following
this does not occur
for $J \approx t$, i.e., in the parameter range relevant to high-$T_C$
superconductivity. On the contrary, in current
applications$^{3,4}$ the simple \tJmb is generally
discussed in the range of intermediate and large $J/t$, where neither
the \tJ nor the \tJt models have a firm physical basis.

The problem of interacting electrons can be faced from
a different viewpoint looking for interactions
not described by standard \hum s. The possible role of
\orb \reb in favoring superconductivity has recently been
emphasized by several authors$^{7-11}$.
Orbital \reb can be modeled in terms of a \humb
with occupation-dependent charge-transfer integrals.
In the present paper it will be shown that in
a physically relevant parameter region the \humb with
occupation dependent hopping
reduces to a \tJtmb with arbitrarily large $J$ values.
The equivalence of the two models, and their differences
in respect of both $t-U$ and \tJm s, are tested by
comparing the ground state energy and the charge correlation
functions obtained by exact diagonalization of a 12-site
one-dimensional chain with periodic boundary conditions.

The \humb with occupation dependent hopping
is defined by the \ha :

$$\eqalign{
&H_{t_1-t_2-t_3-U} =
\cr &
- t_1 \sum_{<i,j>, \sigma}
( a^+_{i, \sigma} a_{j, \sigma} + {\rm h.c.})
(1 - n_{i, - \sigma}) (1 - n_{j, - \sigma})
\cr &
- t_2 \sum_{<i,j>, \sigma}
( a^+_{i, \sigma} a_{j, \sigma} + {\rm h.c.})
[ n_{i, - \sigma} (1 - n_{j, - \sigma}) +
\cr &
\phantom{- t_2 \sum_{<i,j>, \sigma}
( a^+_{i, \sigma} a_{j, \sigma} + {\rm h.c.})}
(1 - n_{i, - \sigma}) n_{j, - \sigma} ]
\cr &
- t_3 \sum_{<i,j>, \sigma}
( a^+_{i, \sigma} a_{j, \sigma} + {\rm h.c.})
n_{i, - \sigma} n_{j, - \sigma} +
U \sum_i n_{i, \alpha} n_{i, \beta}
}\eqno(5)$$

This \hab has been recently proposed$^{7-11}$,
within the Hubbard scheme,
to account for the orbital \reb effect,
that is, the \reb of a \dob orbital with
respect to a \sib one.
This effect originates different $t$-values
for the three possible hopping processes, depending on the
electron occupation of the two sites involved
in the hopping. A rigorous, first-principle derivation of the
\timb is reported in$^{10,11}$: the assumption that the
orbital \reb is orthogonal to the bond direction implies$^{10}$
the ordering $t_1 > t_2 > t_3$, whereas, if one allows
the \dob orbital to expand also in the bond direction$^{11}$,
one gets truly independent $t_1 , t_2 , t_3$ values.
Several "limiting cases" for the $t_1 : t_2 : t_3$ ratio
can be considered, originating different physical properties:

%\item{(1:1:1)} the standard \huh
{(1:1:1)} the standard \huh

%\item{(1:$1\over 2$:0)} the so-called $t-U-X$ model for
{(1:$1\over 2$:0)} the so-called $t-U-X$ model for
$X = - {1 \over 2} t$, extensively discussed by Hirsch$^{7}$

%\item{(1:0:0)} a model which presents an interesting phase
{(1:0:0)} a model which presents an interesting phase
diagram, discussed in$^{11}$

%\item{(0:1:0)} the case of interest in the present context
{(0:1:0)} the case of interest in the present context

A thorough discussion of the physical origin of the different
$t_1 : t_2 : t_3$ ratios is postponed to a forthcoming
paper$^{11}$, together with a detailed description of
the properties of the corresponding models.
Here we concentrate attention on the last case.

As it will be shown in detail in$^{11}$,
in order to get small $t_1/t_2$ and $t_3/t_2$ ratios, two
conditions have to be fulfilled:
(a) the presence of a {\it large \orb \re}, both
perpendicularly and along the bond direction,
in such a way that the overlap between nearest neighbor
non-relaxed \or s is negligible, whereas the overlap
between relaxed (i.e., \do) \or s is "large" ($\approx
0.2$); (b) a concurrent
large \reb of the core \or s (note
that all occupied \or s not directly involved in the electron
transfer have to be intended as "core shells").
We believe that both conditions are fulfilled
in models describing the motion of holes in the $O^=$
sublattice in CuO planes. In fact a large \reb of the
oxygen $p-$\or s (both directly involved in the hopping and
"spectator") is expected following the
variation of the oxygen charge from $-1$ to $-2$$^{10,11}$.

In the large-$U$ limit, the relaxed orbital \humb with
$(t_1 : t_2 : t_3) = (0:1:0) $ can be reduced, via a standard
second-order perturbative expansion which eliminates \dob sites$^{2}$,
to a \tJtmb with $J = (4 t_2^2 / U)$,
$t^\prime = (t_2^2 / U)$ and $t = 0$.
The derivation holds true also for finite but
small $t_1 / t_2$ or $t_3 / t_2$ ratios: in this case,
setting for the sake of simplicity $t_1 = t_3$, one gets the
same \tJtmb as before, but with $t = t_1$.
In the large-$U$ limit, the \timb
is therefore equivalent to a \tJtmb with
$J/t = (4t_2^2/t_1 U) = (4t_2/U) (t_2 / t_1)$.
With respect to the $J/t$ ratio derived from the
large-$U$ expansion of the standard \humb
($J/t = (4 t / U)$), one gets a further factor
$(t_2 / t_1)$, which can compensate the term $(4 t_2 / U)$
and give a $J/t$ ratio of order 1 or even larger.

In order to confirm these predictions exact diagonalization studies
(diagrammatic valence bond$^{12}$) have been
performed on 3/4-filled finite-size systems.
This band-filling is particularly interesting:
organic superconductors are in fact 3/4-filled and the highest
$T_C$ for CuO superconductors are observed for filling in
the 9/10 $\div$ 3/4 range$^{13}$. Moreover, recent numerical
calculations on the \tJm $^{14}$ show that
superconducting correlations are enhanced near 3/4 filling
for both 1- and 2-dimensional systems.
In Fig.1 we report as a function of $J/t$
the ground state energy calculated for
a 12-site linear chain with periodic boundary conditions and
total number of electrons $n=6$ (\tJ -like models,
\dob sites are strictly forbidden) or $n=18$
($t-U$-like models, in this case $E - 6*U$ is
actually plotted).
Fig.1, top panel, clearly shows that \tJt and \tJmb
are different, apart from a very narrow region of small
$J$ (for $J$=1 the energy difference between the two models
is larger than $10\%$, and for $J$=2 it is larger than $20\%$).
Even worst is the comparison between \tJt and $t-U$ models,
whereas \tJ and $t-U$ models have similar energy
(differences smaller than $10\%$) up to $J \approx 2$ ($U=2$).
In Fig.1, bottom panel, we compare the energy calculated
for the \tJtmb with that calculated for the \timb
for fixed $U/t_2$ values and variable $t_1/t_2$ ratios.
For large $U/t_2$ ($U=10 t_2$), the agreement between the two models
is quantitative, even at very large $J$ values ($J/t$=16):
the energy difference is smaller than $7\%$.
At intermediate values of $U$ ($U=5 t_2$), the agreement is worse,
still the energy differences stay lower than $20\%$ up
to $J=16$.
Finally at small $U$ ($U= t_2$) the large-$U$ expansion of the
\timb is no more possible.

A sounder comparison among the models involves ground state
correlation functions.
For 3/4-filled systems three extreme phases
can be envisaged$^{15}$: a monomer liquid
(ML) phase where ${\cal D}$- (\do) and ${\cal S}$- (\si ) sites
regularly alternate, a dimer liquid (DL) phase where pairs of
${\cal D}$-sites alternate with pairs of ${\cal S}$-sites,
and a phase segregated (PS) state.
Recent numerical calculations
on the $t-J-V$ model$^{14}$ have confirmed the original
suggestion by Kivelson {\it et al.}$^{15}$ that superconductivity
is enhanced in
regions where DL phase dominates. In DL phase in fact
the electrons cluster but do not condense.
Information on the relative stability of the three phases
is offered by charge correlation functions $O_d$
defined as:

$$
O_d = {2 \over N} ( < \sum_i n_i n_{i+d} > - m)
\eqno(6)$$
where the brackets indicate the ground state expectation value
and $m$ is set to 5 for $t-U$-like models ($n=18$),
and to 1 for \tJ -like models ($n=6$).
The stability of ML, DL and PS
phases increases with increasing $O_{d=1}$, $O_{d=2}$, $O_{d=6}$,
respectively.

The $O_d$-vs-$J$ curves, calculated for the 12-site chain
with different models are reported in Fig.2.
The \tJ and $t-U$ models show a qualitatively different behavior from
\tJtm . On the other hand the \timb for all $U$ values
gives a phase diagram similar to the \tJtm , showing at small $J$
a region of stability for ML phase, and, at large $J$,
a region of stability for DL phase.
A quantitative comparison of the models gives results similar
to those already obtained from the ground state energy, with
the largest differences in correlation functions between
\tJt and \tim s being of the order of $7\%$ for $U=10$ and
$17\%$ for $U=5$, much larger ($\approx 30\%$) for $U=1$
(note however that the behavior
of \timb with small $U$ is in any case very interesting).

In summary, the large-$U$ expansion of the \timb originates
a \tJtmb with arbitrarily large $J/t$ ratios.
The expansion is quantitatively correct for large $U$
($U/t_2$=10) and semi-quantitatively correct for intermediate $U$
($U/t_2$=5).
On the other hand, $t-U$ model can never originate \tJtmb
with large $J$. We have also demonstrated that \tJtmb reduces
to the \tJmb only at small values of $J/t$.
We finally observe that, for what concerns superconductivity,
$t-U$ model appears not very interesting, showing
a dominating ML phase at all $U$ values, in agreement with
the exact solution of the \humb in the one-dimensional case$^{16}$.
\tJmb exhibits a more interesting behavior, but the
region where DL phase dominates is very narrow since
PS phase becomes largely dominating with increasing $J$.
On the contrary, in the \tJt \hab the $t^\prime$ term
(a second-neighbor
{\it hopping} term proportional to $J$) preserves the
systems from collapsing onto phase segregation.
For the \tJtmb at large $J$ and the \timb for small $t_1/t_2$
and any $U/t_2$ value,
therefore, DL phase dominates, thus suggesting a possible role
of these models in the theoretical interpretation of
high-T$_C$ superconductivity
(note that the (0:1:0) range of parameters has been recently$^{17}$
suggested, from a different viewpoint,
to be favorable to superconductivity).

\vskip 1truepc

\parindent=0cm
{\it Acknowledgements} -
We thank A. Girlando and J. Voit for useful discussions.
This research has been performed within the framework of
"Progetto Finalizzato Materiali Speciali per Tecnologie
Avanzate" of Consiglio Nazionale delle Ricerche (Italy).
Financial support from the Italian MURST is also acknowledged.
\parindent=0.9truecm

\newpage

{\bf References }

\item{1} P. W. Anderson,
{\it Science} {\bf 235} (1987) 1196.

\item{2} P. Fulde, {\it Electron Correlations in Molecules and Solids},
Springer Series in Solid State Sciences 100 (Springer, Berlin, 1991)
pp. 288-290.

\item{3} E. Dagotto,
{\it Int. J. Modern Phys. B} {\bf 5} (1991) 77, and references therein.

\item{4} M. Ogata, M. U. Luchini, S. Sorella and F. F. Assaad,
{\it Phys. Rev. Lett.} {\bf 66} (1991) 2388;
F. F. Assaad and D. W\"urtz,
{\it Phys. Rev. B} {\bf 44} (1991) 2681;
M. Imada and Y. Hatsugai,
{\it J. Phys. Soc. Japan} {\bf 58} (1989) 3752.

\item{5} B. Friedman, X. Y. Chen and W. P. Su,
{\it Phys. Rev. B} {\bf 40} (1989) 4431.

\item{6} J. A. Riera and A. P. Young,
{\it Phys. Rev. B} {\bf 39} (1989) 9697;
K. J. vonSzczepanski, P. Horsch, W. Stephan and M. Ziegler,
{\it Phys. Rev. B} {\bf 41} (1990) 2017;
H. Fehske, V. Waas, H. R\"oder and H. B\"uttner,
{\it Solid State Commun.} {\bf 76} (1990) 1333.

\item{7} J. E. Hirsch and F. Marsiglio,
{\it Phys. Lett. A} {\bf 140} (1989) 122;
J. E. Hirsch, {\it Phys. Rev. B} {\bf 47} (1993) 5351
and references therein.

\item{8} I. O. Kulik, {\it Sov. Superconductivity Phys. Chem. Tech.}
{\bf 2} (1989) 201.

\item{9} J. E. Hirsch,
{\it Phys. Rev. B} {\bf 43} (1991) 11400.

\item{10} A. Fortunelli and A. Painelli, {\it Chem. Phys. Lett.},
(in press).

\item{11} A. Fortunelli and A. Painelli,
(in preparation).

\item{12} S. Ramasesha and Z. G. Soos,
{\it J. Chem. Phys.} {\bf 80} (1984) 3278;
{\it Int. J. Quantum Chem.} {\bf 25} (1984) 1003.

\item{13} H. Zhang and H. Sato,
{\it Phys. Rev. Lett.} {\bf 70} (1993) 1697.

\item{14} E. Dagotto and J. Riera,
{\it Phys. Rev. B} {\bf 46} (1992) 12084;
{\it Phys. Rev. Lett.} {\bf 70} (1993) 682.

\item{15} S. A. Kivelson, V. J. Emery and H. Q. Lin,
{\it Phys. Rev. B} {\bf 42} (1990) 6523.

\item{16} E. H. Lieb and F. Y. Wu,
{\it Phys. Rev. Lett.} {\bf 20} (1968) 1445.

\item{17} A. Ferretti, I. O. Kulik and A. Lami,
{\it Phys. Rev. B} {\bf 47} (1993) 12235.

\newpage

\leftline  {\bf Figure Captions} \par
\vskip 1truepc
\vskip 1truepc
\parindent=0truecm

{Fig.1} The ground state energy ($E$) vs $J$ calculated for
a 12-site chain with periodic boundary conditions and $n=6$
electrons (\tJ and \tJtm s) or $n=18$ ($t-U$ and \tim s;
in this case $E - 6*U$ is actually plotted). Both $E$ and
$J$ are in units of $t$ (for \timb, $J/t=4t_2^2/U t_1$).
In both panels continuous lines refer to \tJtm .
In top panel long- and short-dashed lines refer to \tJ and
$t-U$ models, respectively. In bottom panel
long- and short-dashed and dot-dashed lines refer to \timb
with $U/t_2 = 10, 5, 1$, respectively.

\vskip 1truepc

{Fig.2} Charge correlation functions ($O_d$)
calculated for the 12-site periodic chain with $n=6$ electrons
(\tJ and \tJtm s) or $n=18$ ($t-U$ and \tim s).
On the $x$-axys $J$ is measured in $t$-units ($J/t =
4 t_2^2 / U t_1$ for \tim ).
Continuous, short- and long-dashed lines refer to
the correlation functions relevant to ML, DL, PS
phases ($d=1, 2, 6$), respectively.
Left panels refer to \tim s with $U/t_2$ ratios
specified in the figure.
In the right middle panel (\tJm)
$y$-axys actually ranges from 0. to 1.2.

\parindent=0.9truecm

\end